\documentclass[a4paper]{article}
\usepackage{INTERSPEECH2021,amsmath,graphicx}

\usepackage{multirow}
\usepackage{makecell}
\usepackage[flushleft]{threeparttable}
\usepackage{hyperref}
\usepackage{amsmath}
\usepackage{soul,color}
\usepackage{subcaption}
\usepackage{enumitem}
\usepackage{xcolor}
\usepackage{booktabs}

\title{TalkNet 2: Non-Autoregressive Depth-Wise Separable Convolutional Model for Speech Synthesis with Explicit Pitch and Duration  Prediction}

\name{
Stanislav Beliaev, Boris Ginsburg\thanks{Preprint. Accepted to INTERSPEECH 2021}
}
\address{NVIDIA, Santa Clara} \email{\{stanislavv,bginsburg\}@nvidia.com} 

\begin{document}
\maketitle

\begin{abstract}
We propose TalkNet, a non-autoregressive convolutional neural model for speech synthesis with explicit pitch and duration prediction. The model consists of three feed-forward convolutional networks. The first network predicts grapheme durations. An input text is then expanded by repeating each symbol according to the predicted duration. The second network predicts pitch value for every mel frame. The third network generates a mel-spectrogram from the expanded text conditioned on predicted pitch. All networks are based on 1D depth-wise separable convolutional architecture.
The explicit duration prediction eliminates word skipping and repeating. The quality of the generated speech  nearly matches the best auto-regressive models --
TalkNet trained on the LJSpeech dataset got a MOS of $4.08$. The model has only 13.2M parameters, almost $2$x less than the present state-of-the-art text-to-speech models. The non-autoregressive architecture allows for fast training and inference.
The small model size and fast inference make TalkNet an attractive candidate for embedded speech synthesis. 
\end{abstract}


\section{Introduction}

Neural Network (NN) based models for text-to-speech (TTS) have outperformed both concatenative and statistical parametric speech synthesis in terms of speech quality. 
Neural TTS systems typically have two stages. In the first stage, a model generates mel-spectrograms from text. In the second stage, a NN-based vocoder synthesizes speech from  mel-spectrograms. Most NN-based TTS models have an encoder-attention-decoder architecture \cite{bahdanau2014neural}, which has been observed to have some common problems \cite{Fastspeech2019}:
\begin{enumerate}
    \item A tendency to repeat or skip words due to attention failures. To handle this issue,  one can use additional mechanisms to encourage monotonic attention \cite{Tacotron2,DeepVoice3}.
    \item Slow inference relative to parametric models.
    \item No easy way to control of word duration or voice speed.
\end{enumerate}
\begin{figure}[!ht]
  \centering
  \includegraphics[width=0.95\linewidth]{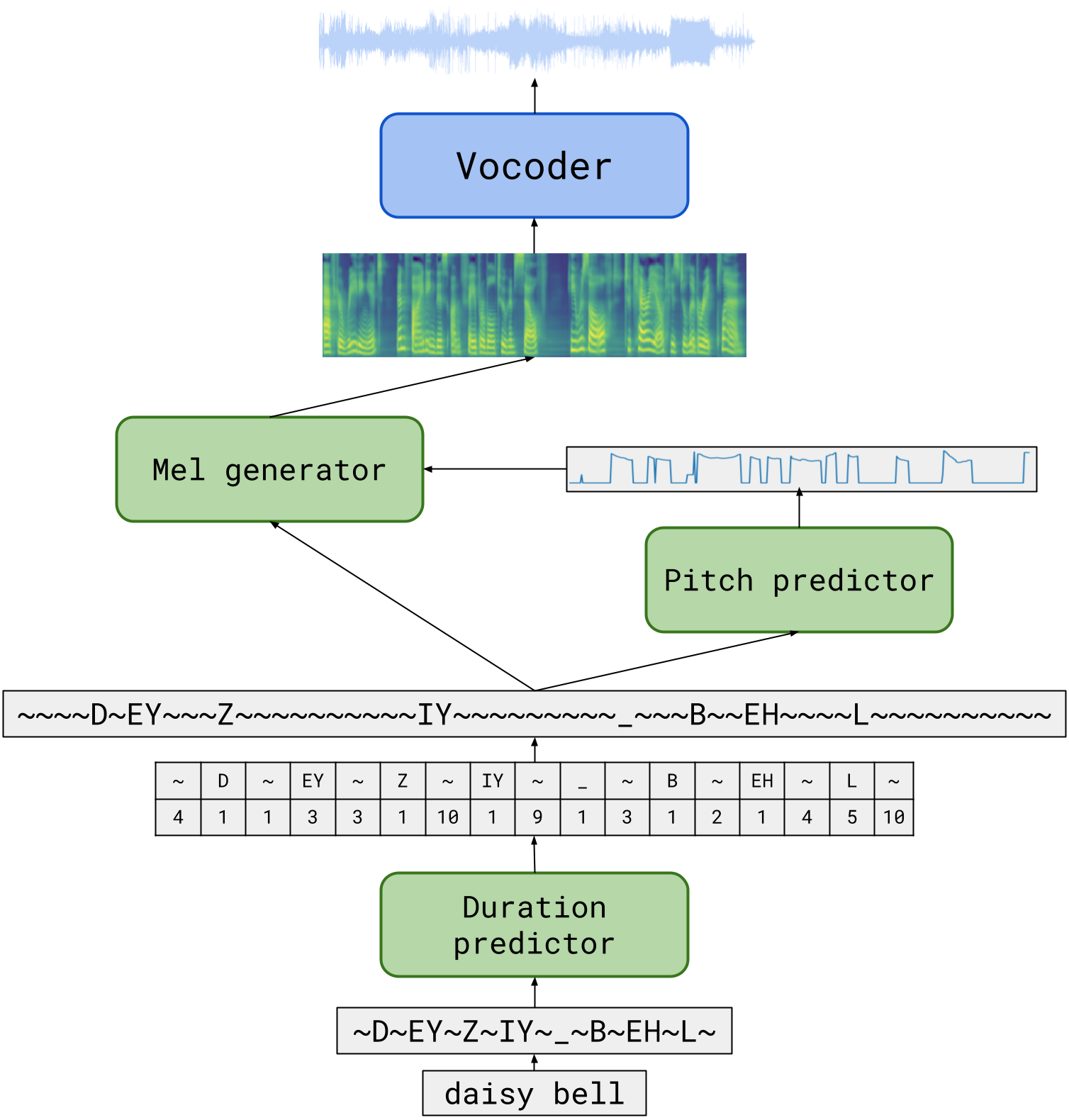}
  \caption{TalkNet converts text to speech, using a grapheme duration predictor, pitch predictor and a mel-spectrogram generator. We use $\sim$ to denote the blank symbol.}
  \label{fig:architecture}
\end{figure}
We propose a new neural TTS model to address these three challenges. The model consists of three convolutional networks.
The first network predicts grapheme durations. The explicit duration predictor replaces attention-based alignment to prevent word skipping and repeating. Next we expand an input text based on the grapheme predicted duration. The second network predicts pitch along expanded sequence. The third network generates mel-spectrograms from an expanded text and predicted pitch. Finally, we use an external vocoder - WaveGlow~\cite{Waveglow2019} or HiFi-GAN vocoder~\cite{hifigan} - to synthesize audio from mel-spectrograms (see Figure~\ref{fig:architecture}).

To train the grapheme duration predictor, we use  the ground truth alignment between input characters and speech. To  obtain this alignment, we proposed a new method which uses an external automatic speech recognition (ASR) Connectionist Temporal Classification (CTC) based model.
Instead of using the  greedy output of ASR model like in DeepVoice~\cite{DeepVoice1}, we extract the alignment by calculating the most probable path in the lattice using the Viterbi algorithm.

The quality of speech generated by TalkNet combined with a HiFi-GAN vocoder nearly matches the best auto-regressive models: TalkNet trained on the LJSpeech got a MOS of $4.08$. The TalkNet architecture is based on 1D depth-wise separable convolutions, similar to QuartzNet \cite{quartznet}. As a result, TalkNet has only 13.2M parameters - almost 2X less than models with similar quality, e.g. the Transformer-based FastSpeech~\cite{Fastspeech2019, fastspeech2}, or  FastPitch~\cite{2021fastpitch}. 
The non-autoregressive architecture enables fast parallel training and inference comparing to auto-regressive models like Tacotron 2~\cite{Tacotron2}.
The TalkNet inference is $132$ times faster than real-time with batch size 1, and $~2531$ times faster for batch size 32.

The high quality of generated speech combined with low compute and memory requirements make the TalkNet an attractive candidate for embedded speech synthesis. 

\section{Related work}
A traditional parametric TTS pipeline has the following stages: grapheme-to-phoneme conversion, a phoneme duration predictor, an acoustic frame-level feature generator, and a vocoder \cite{Taylor2009}. Neural networks for TTS have been explored in 1990 \cite{Weijters1993, Tuerk1993, Karaali1996, Karaali1998} but the speech quality was below the quality of traditional methods at that  time. Deep learning re-introduced NNs to TTS in 2015: Zen et al \cite{Zen2015, Zen2016} proposed a hybrid NN-parametric TTS model, where deep NNs are used to predict the phoneme duration and to generate frame-level acoustic features.
Merlin TTS system \cite{Merlin2016} supports various NN architectures, including a standard feed-forward NNs, and recurrent neural networks (RNNs) to predict acoustic features, which are then passed to a traditional vocoder to produce the speech. 

DeepVoice models~\cite{DeepVoice1, DeepVoice2} adopted the traditional TTS pipeline, but they replace all components with NNs. To train the phoneme duration predictor, an auxiliary CTC-based model for phonetic segmentation was used to annotate  data with phoneme boundaries. 
Tacotron 2~\cite{Tacotron1, Tacotron2} one of the first end-to-end neural TTS model which are based on  encoder-attention-decoder architecture. The encoder is composed of three convolutional layers plus a single bidirectional LSTM. The decoder is a  RNN with location-sensitive monotonic attention.

The sequential nature of RNN-based models limits the training and inference efficiency. DeepVoice 3 \cite{DeepVoice3} replaces an RNN with a fully-convolutional encoder-decoder with monotonic attention. Switching from RNN to a convolutional neural network (CNN) makes training faster, but the  inference is still auto-regressive. ParaNet neural TTS model also does not use RNNs \cite{Paranet2019}. ParaNet is a convolutional encoder-decoder with attention. It distills attention from a teacher auto-regressive TTS model. Lastly, Transformer-TTS \cite{TransformerTTS} replaces an RNN-based encoder-decoder with a Transformer-like \cite{vaswani2017attention}  architecture.

Attention-based models (e.g.  Tacotron, Transformer-TTS, ParaNet etc) occasionally miss or repeat words \cite{Paranet2019}.
To prevent word skipping and repeating, FastSpeech~\cite{Fastspeech2019} proposes a feed-forward Transformer-based model with  explicit length regulator. FastSpeech expands the sequence of phonemes according to a predicted duration in order to match the length of a mel-spectrogram sequence. The target phoneme duration is extracted from the attention alignment in an external pre-trained Tacotron 2 model.  However, FastSpeech had several disadvantages: 1) the target mel-spectrograms distilled from teacher model suffer from information loss, 2) the phoneme duration extracted from the attention in teacher model is not accurate enough. 
FastPitch~\cite{2021fastpitch} is a logical extension of the FastSpeech model with explicit pitch modeling. Pitch is extracted and averaged over each grapheme duration, which helps make prediction process easier at the inference stage. FastPitch also relies on a pretrained attention TTS model to extract durations and it archives a notable quality increase in terms of Mean Opinion Socre (MOS).

\section{System architecture}

TalkNet splits the text-to-spectrogram generation into three separate modules (see Figure~\ref{fig:architecture}). The first module, the duration predictor, aligns input graphemes in time with respect to the audio features. The second module, the pitch predictor, reconstructs pitch along expanded input tokens. The third module generates  mel-spectrograms from time-aligned input sequence and pitch pattern. We use feed-forward CNNs for all modules, so both training and inference are non-autoregressive. 

\subsection{Grapheme duration predictor}
\begin{figure}[!ht]
  \centering
  \includegraphics[width=0.95\linewidth]{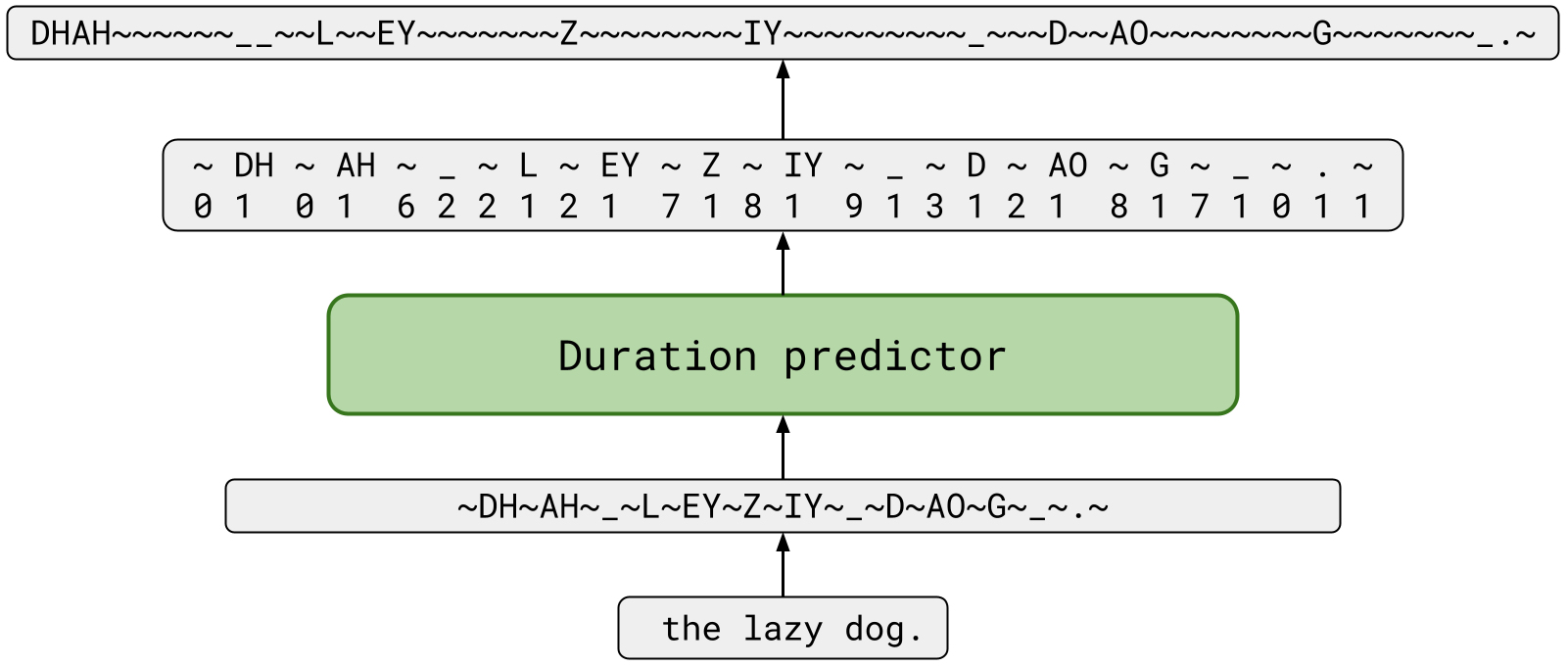}
  \caption{Grapheme duration prediction.}
  \label{fig:durs-logic}
\end{figure}

This model predicts the length of the mel-spectrogram part corresponding to each grapheme in the input including punctuation. First, the grapheme duration predictor inserts a blank symbol $\sim$ between every two input tokens. 
Next, it predicts the duration for each input token including blank. Finally, the sequence of input graphemes is expanded by repeating each token according to the predicted duration (see Figure~\ref{fig:durs-logic}).

\begin{figure}[ht!]
  \centering
    \caption{Basic QuartzNet block. Both the grapheme duration predictor and the mel-spectrogram generator are 1D time-channel convolutional networks based on QuartzNet \cite{quartznet}.}
  \includegraphics[width=0.45\linewidth]{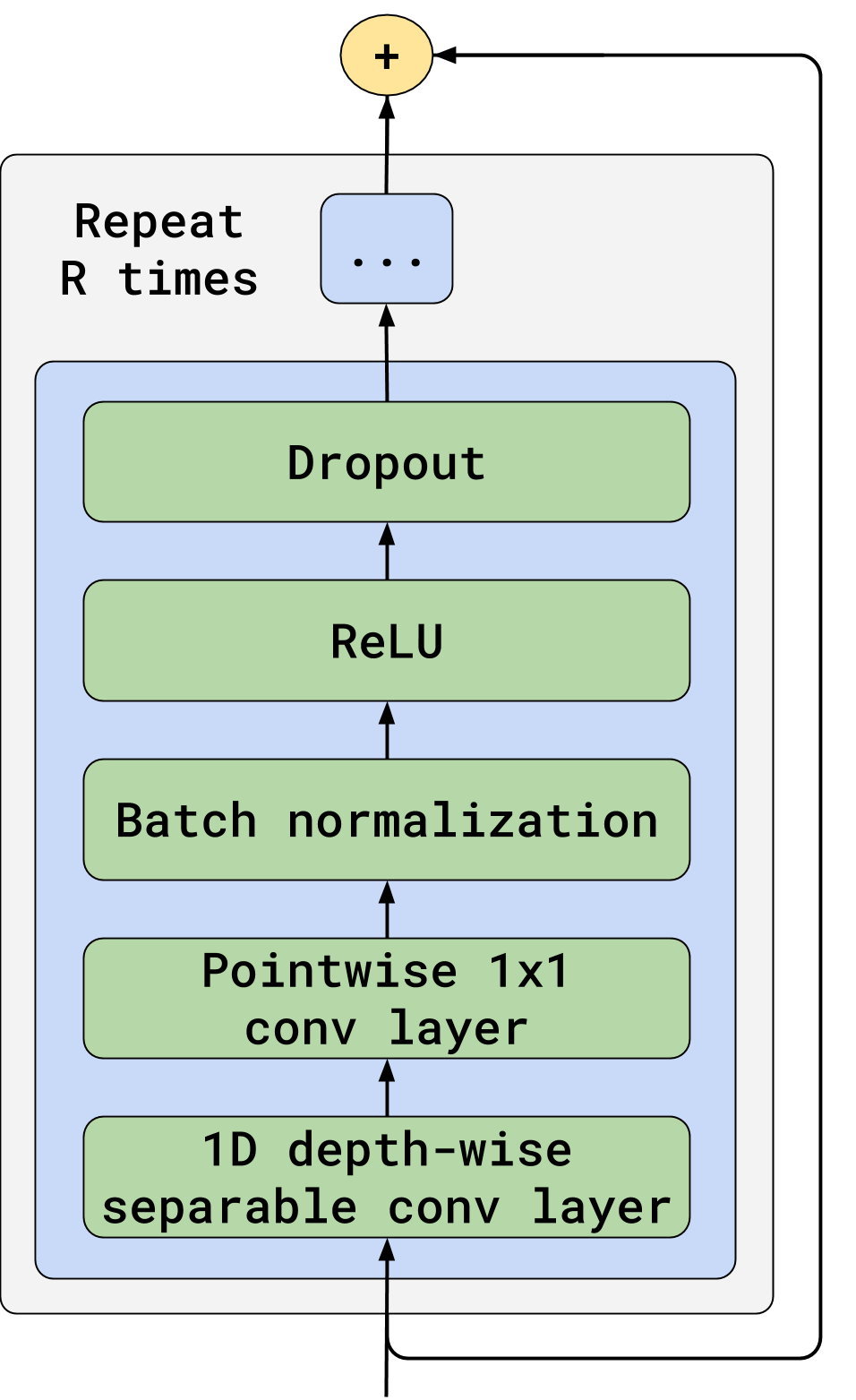}
  \label{fig:quartznet_basic_block}
\end{figure}

The grapheme duration predictor is a 1D time channel separable convolutional NN based on the QuartzNet architecture~\cite{quartznet}. The model has five residual blocks with five sub-blocks per block. A sub-block consists of a 1D time-channel separable convolution, a $1\times 1$ pointwise convolution, batch normalization, ReLU, and dropout (see Figure~\ref{fig:quartznet_basic_block}). 
There are two additional layers: the grapheme embedding layer, and a $1\times1$ convolution before the loss (see Table~\ref{tab:durs-model}).

\begin{table}[ht!]
\caption{Both grapheme duration and pitch predictors has identical architectures based on QuartzNet-5x5.}
\label{tab:durs-model}
\centering
\scalebox{0.8}{
\begin{tabular}{c c c c c} 
\toprule
\textbf{Block} &
\textbf{\thead{\# Sub\\Blocks}} &
\textbf{\thead{\# Output\\Channels}} &
\textbf{Kernel Size} &
\textbf{Dropout} \\
\midrule
Embed & 1 & 64  & 1 & 0.0  \\
Conv1 & 3 & 256 & 3 & 0.1  \\
$B_1\dots B_5$ & 5 & 256 & \{5, 7, 9, 11,13\} & 0.1  \\
Conv2 & 1 & 512 & 1 & 0.1  \\
Conv3 & 1 & $32$ & 1 & 0.0 \\
\midrule
\textbf{Parameters, M} & & & & \textbf{2.3} \\
\bottomrule
\end{tabular}
}
\end{table}



\begin{figure}[!ht]
\centering
\includegraphics[width=.48\linewidth]{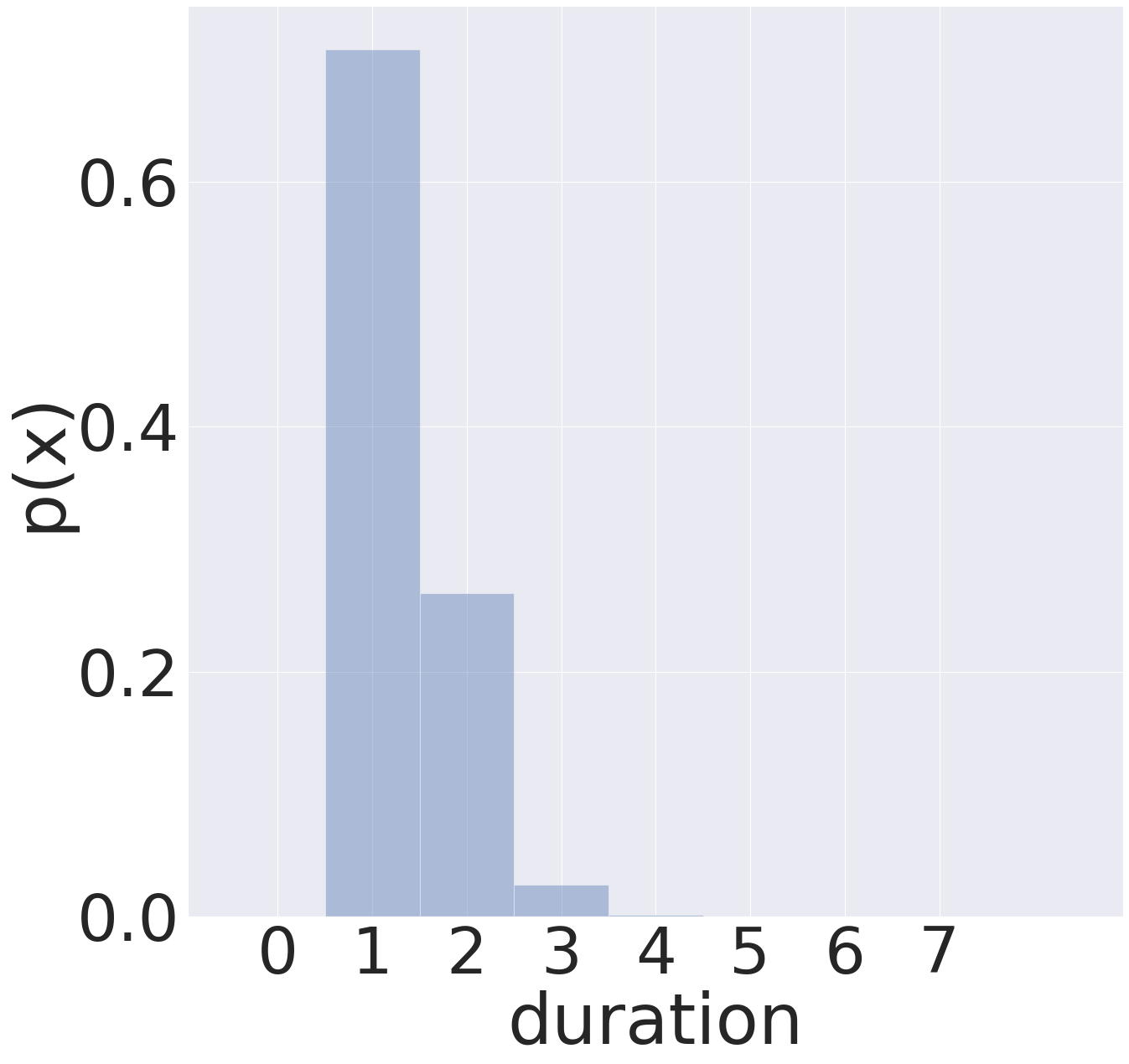}
\includegraphics[width=.48\linewidth]{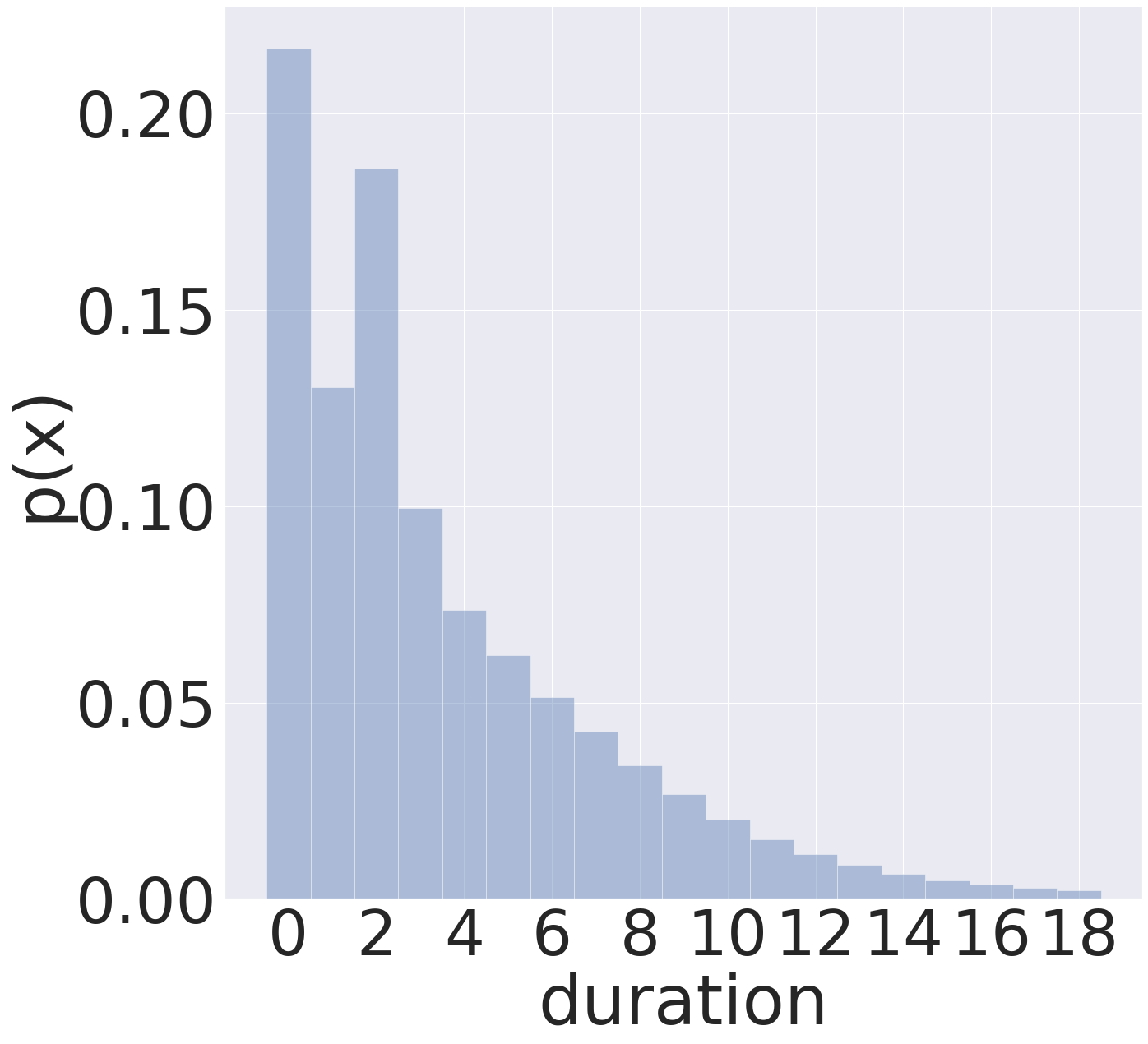}
\caption{The duration distribution for graphemes (left) and for blanks (right) based on CTC output for the LJSpeech dataset.}
\label{fig:durs-dist}
\end{figure}

We train a duration predictor using $L_2$ loss with logarithmic targets, similar to \cite{Fastspeech2019}. We used a log scale for large durations since the grapheme duration distribution has a long tail (see Figure~\ref{fig:durs-dist}). 
We also tried cross-entropy loss with each class corresponding to the character duration, but we found that  speech generated with $L_2$ loss received a slightly higher MOS in our evaluation studies.

\subsection{Pitch predictor}

\begin{figure}[!ht]
  \centering
  \includegraphics[width=0.95\linewidth]{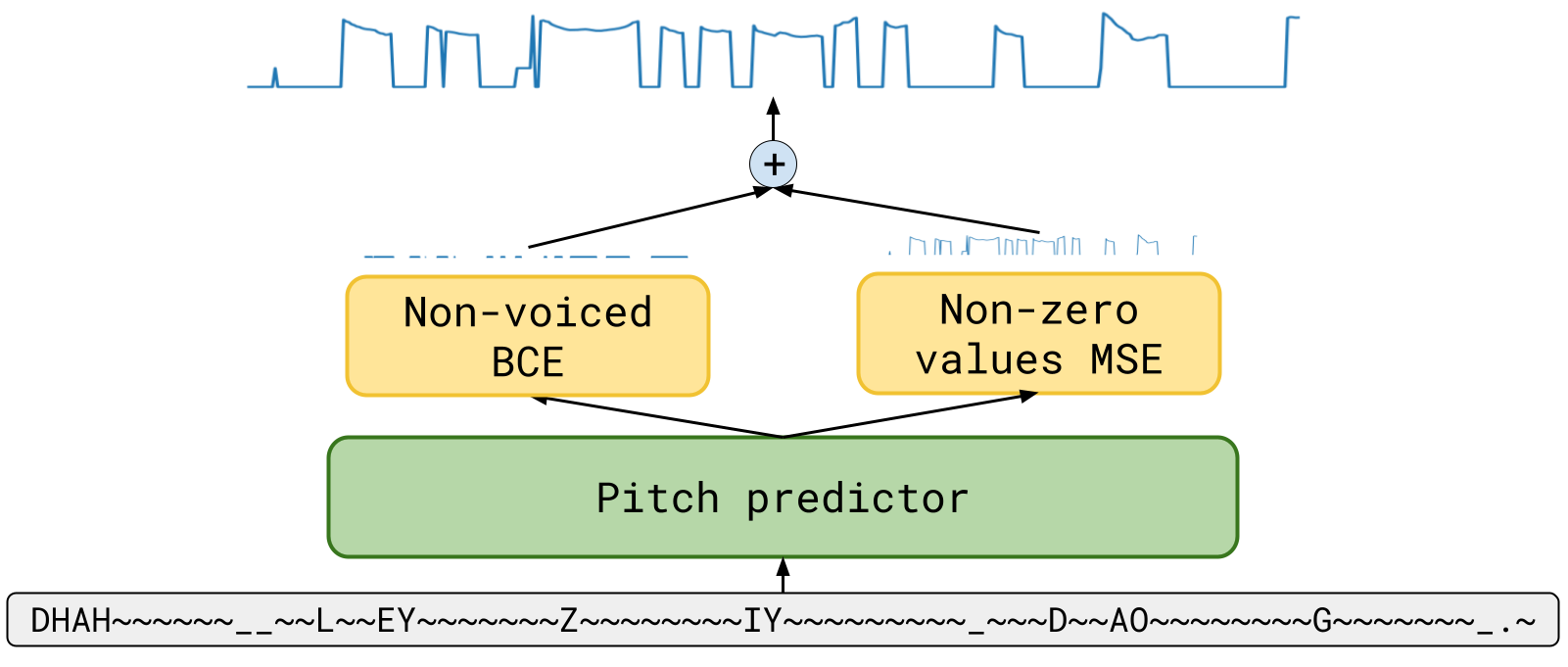}
  \caption{Pitch predictor loss is a sum of two heads: one head is binary cross-entropy (BCE) for non-voiced prediction, another head is mean squared error (MSE) for non-zero values.}
  \label{fig:pitch-logic}
\end{figure}

The pitch prediction network is QuartzNet-5x5, similar to the duration predictor (see Table \ref{tab:durs-model}). The model has two heads: one head is binary cross-entropy (BCE) for non-voiced prediction, another -- mean squared error (MSE) for non-zero values. Final loss is computed as sum of two head's losses with equal weights coefficients. We scale MSE loss with global mean and standard deviation statistics to make it comparable to BCE  (Figure~\ref{fig:pitch-logic}).
To train the pitch predictor we used as input the sequence of text tokens expanded with ground truth token durations.  Following~\cite{shen2020nonattentive}, we use  Gaussian embedding layer for tokens with normal distribution proportional to the token duration. 

\subsection{Mel-spectrogram generator}
\begin{table}[ht!]
\caption{Mel-spectrogram generator is based on QuartzNet-9x5.}
\label{tab:mel-predictor}
\centering
\scalebox{0.85}{
\begin{tabular}{c c c c c} 
 \toprule
  \textbf{Block} &
  \textbf{\thead{\# Sub\\Blocks}} &
  \textbf{\thead{\# Output\\Channels}} &
  \textbf{Kernel Size} &
  \textbf{Dropout} \\
 \midrule
Embed & 1 & 256 & 1 & 0.0 \\
Conv1 & 3 & 256 & 3 & 0.1 \\
$B_1\dots B_6$ & 5 & 256 & \{5,7,9,13,15,17\} & 0.1  \\
$B_7\dots B_9$ & 5 & 512 & \{21,23,25\} & 0.1  \\
Conv2 & 1 & 1024 & 1 & 0.1 \\
Conv3 & 1 & 80 & 1 & 0.0 \\
 \midrule
 \textbf{Parameters, M} & & & & \textbf{8.5} \\
 \bottomrule
\end{tabular}
}
\end{table}
The mel-spectrogram generator is QuartzNet-9x5 with a mean squared error (MSE) loss, see Table~\ref{tab:mel-predictor}.
Mel-spectrogram generator is conditioned on pitch values through residual connection with a simple linear resize layer. We also use the Gaussian embedding for expanded input sequence as in pitch predictor.

\section{Training}

\subsection{Dataset}
The model was trained on LJSpeech dataset~\cite{ljspeech} which was split into three sets: $12,500$ samples for training, $100$ samples for validation, and $500$ samples for testing. The text was lower-cased while leaving all punctuation intact.
We convert ground truth audio to mel-spectrograms using a Short-Time Fourier Transform (STFT) with 50 ms Hann window and 12.5 ms frame hop. Ground truth pitch was extracted using \textit{pysptk} library~\cite{pysptk} with values aligned with mel-spectrogram frames.

\subsection{Ground truth duration extraction}
The non-autoregressive model rely on audio/text alignment based on the prediction of duration for each input token. To obtain the ground truth durations one can use an external aligner, for example Montreal Forced Aligner \cite{mfa2017}. We extracted the ground truth alignment from a CTC ASR model, similar to \cite{DeepVoice1}.
The CTC model assigns a probability to each of graphemes from the phonetic alphabet plus an auxiliary blank symbol $\sim$. The blank symbol acts as an intermediate state between two neighbouring graphemes, and its duration roughly corresponds to the length of the transition from one token to another. 

We train ASR model with text with punctuation, converted to phonemes using \textit{g2p} seq2seq model~\cite{g2pE2019}.
We use QuartzNet-5x5~\cite{quartznet} with a minor modification: we set the stride in the first convolutional layer to 1 to make the length of the CTC input equal to the mel-spectrogram. Model was trained on LibriTTS~\cite{libritts}, and it  achieves a phoneme error rate (PER) of $8.5\%$ on LibriTTS test-clean.

Instead of just using the CTC output which can have errors, we  apply the Viterbi alignment algorithm between ground truth text tokens and NN output with log probabilities. The algorithm finds most probable alignment by estimating maximum joint probability for prefix of ground truth input tokens using dynamic programming. The algorithm has two passes: forward pass -- to calculate  maximum sum of output log probabilities over time, and backward pass -- to traverse best path in reverse mode and count number of steps we stays at each grapheme. The overall algorithm similar to CTC alignment, but instead of summing over all correct paths, we use only the most probable path.  

\subsection{Grapheme duration and pitch predictors training}

The NN for both grapheme duration and pitch was trained using the Adam with $\beta_1=0.9,\beta_2=0.999,\epsilon=10^{-8}$, a weight decay of ${10}^{-6}$ and gradient norm clipping of $1.0$. We used a cosine decay learning rate policy starting from $10^{-3}$ and decreasing to $10^{-5}$ with a $2\%$ warmup. We used a batch of 256 for one 16GB GPU and scaled learning rate for multi-GPU setups. Training duration predictor for $200$ epochs takes $\approx 1.3$ hours on one V100 GPU, and 11 minutes on a DGX1 server with 8 GPUs in mixed precision~\cite{micikevicius2017mixed}. Training pitch predictor takes about 2 hours on one V100 GPU, and 25 minutes on DGX1. The duration predictor accuracy is around $81\%$, but it covers approximately $95\%$ of classes within an absolute distance of $1$.

\subsection{Mel-spectrogram generator training}

We train the mel-spectrogram generator  with the same parameters as above. We used a batch size of 64 for one 16GB GPU, and scale the learning rate for a multi-GPU setup. Training for 200 epochs takes $\approx 8$ hours for one V100 GPU and less than 2 hours for DGX1.

\section{Results}

\subsection{Audio quality}
We conduct the MOS (mean opinion score) evaluation for generated speech using Amazon Mechanical Turk. We compared five types of samples: 1) ground truth audio, 2) ground truth mel-spectrogram converted with HiFi-GAN vocoder, 3) Tacotron 2 + HiFi-GAN, and 4) FastPitch + HiFi-GAN 5) TalkNet + HiFi-GAN. We used NVIDIA's implementation for Tacotron 2 and HiFi-GAN with "all-purpose" checkpoint fine-tuned with output of different models. 
We tested $100$ audio samples with $10$ people per sample. The scores ranged from $1.0$ to $5.0$ with a step of $0.5$. TalkNet speech quality comes quite close to Tacotron 2 and FastPitch (see Table~\ref{tab:mos}).

\begin{table}[!ht]
\centering
\caption{MOS scores with $95\%$ confidence interval}
\label{tab:mos}
\scalebox{1.00}{
    \begin{tabular}{l c} 
    \toprule
    \textbf{Model} &
    \textbf{MOS} \\
    \midrule
    Ground truth speech & $4.32 \pm 0.04$ \\
    Ground truth mel + HiFi-GAN & $4.20 \pm 0.04$ \\
    Tacotron 2 + HiFi-GAN & $4.20 \pm 0.04$ \\
    FastPitch + HiFi-GAN & $4.13 \pm 0.05$ \\
    \midrule
    TalkNet + HiFi-GAN & $4.08 \pm 0.05$ \\
    \bottomrule
    \end{tabular}
}
\end{table}

TalkNet is very robust with respect to missing or repeated words compared to auto-regressive TTS models such as Tacotron 2 or Transformer TTS. We evaluated the  robustness of TalkNet on 50 hard sentences from the FastSpeech paper~\cite{Fastspeech2019} and found that TalkNet practically eliminates missed or repeated words.

\subsection{Inference}

In the inference mode, we first insert blank symbols into the tokenized input text between every two graphemes. The obtained sequence is passed through the grapheme duration predictor. The input sequence is expanded with each token repeated according to the predicted duration. The second network predicts pitch: one head predicts silence parts, another head predicts values for voiced parts corrected with global statistics. The third network generates the mel-spectrogram from the expanded grapheme sequence and predicted pitch values.

We compare TalkNet inference with Tacotron 2 and FastPitch~\cite{2021fastpitch}. To measure the inference speed, we generate mel-spectrograms with a batch size equal to $1$ and $32$ for $100$ samples from the LJSpeech test split. The average mel-spectrogram length is $563.93$ frames. TalkNet inference is significantly faster than Tacotron 2. And it scales well with increasing the batch size (see Table~\ref{tab:tts-models-lats}). Since TalkNet does not use an attention mechanism, the inference speed practically doesn't depend on the input length.


{\renewcommand{\arraystretch}{1.0}
\begin{table}[!ht]
\caption{Inference real-time factor (RTF) for mel-spectrogram generation (without vocoder). The RTF was measured with batch size $1$ and $32$ using a V100 GPU and averaged over 100 test samples from LJSpeech using PyTorch inference with AMP FP16 mode.}
\label{tab:tts-models-lats}
\centering
\scalebox{1.00}{
\begin{tabular}{l r r} 
\toprule
\textbf{Model} & 
\textbf{\thead{RTF,\\Batch size 1}} &
\textbf{\thead{RTF,\\Batch size 32}} \\
\midrule
Tacotron 2~\cite{Tacotron2}  & $13$ & $233$ \\
FastPitch~\cite{2021fastpitch}  & $573$  & $3502$ \\
\midrule
TalkNet & $132$ & $2531$ \\
\bottomrule
\end{tabular}
}
\end{table}
}

\section{Conclusions}

In this paper, we present TalkNet, a non-autoregressive fully convolutional neural speech synthesis  system. The model is composed of three convolutional networks: a grapheme duration predictor, a pitch predictor and a mel-spectrogram generator. The model does not require another text-to-speech model as a teacher. 
The explicit duration predictor practically eliminates skipped or repeated words. TalkNet achieves a comparable level of speech quality to Tacotron 2 and FastPitch. 
The model is very compact. It has only $13.2$M parameters, almost $2$x less than similar neural TTS models: Tacotron-2 has 28.2M, and FastSpeech has 30.1M parameters. Training TalkNet takes only around $2$ hours on a server with 8 V100 GPUs. The parallel mel-spectrogram generation makes the inference significantly faster and optimal for scaling with increasing the batch size. This makes TalkNet a great candidate for using in full-fledged TTS system and with streaming mode.

The model, the training recipes, and audio samples will be released as part of the NeMo toolkit \cite{nemo2019}.


\section{Acknowledgments}
The authors thank Jocelyn Huang, Rafael Valle, Felix Kreuk, Adrian Lancucki, Vitaly Lavrukhin, Jason Li, Christopher Parisien, and Joao Felipe Santos for the helpful feedback and review. 

\bibliography{bibliography}
\end{document}